\documentclass[useAMS,usenatbib]{mn2e}
\usepackage{epsfig}

\title[A proper motion study of the globular cluster M55]{
A proper motion study of the globular cluster M55} 

\author[K. Zloczewski, J. Kaluzny and I. B. Thompson]
{K. Zloczewski$^1$\thanks{E-mail: kzlocz@camk.edu.pl (KZ); 
jka@camk.edu.pl (JK);\newline ian@obs.carnegiescience.edu (IBT);}, J. Kaluzny$^1$ and I. B. Thompson$^2$
\\
$^1$Nicolaus Copernicus Astronomical Center, ul. Bartycka 18, 00-716 Warsaw, Poland\\
$^2$The Observatories of the Carnegie Institution of Washington, 813 Santa Barbara St., Pasadena, CA 91101\\
}

\begin{document}

\date{Accepted 2010 -- --. Received 2010 -- --; in original form 2010 -- --}

\pagerange{\pageref{firstpage}--\pageref{lastpage}} \pubyear{2010}

\maketitle

\label{firstpage}

\begin{abstract}
We have derived the absolute proper motion (PM) of the globular cluster M55
using a large set of CCD images collected with the du Pont telescope between
1997 and 2008. We find 
($\mu_{\alpha}cos\delta$, $\mu_{\delta}$) = ($-$3.31 $\pm$ 0.10, $-$9.14 $\pm$ 0.15) mas/yr 
relative to background galaxies. 
Membership status was determined for 16945 stars with $14<V<21$ 
from the central part of the cluster. 
The PM catalogue includes 52 variables of which 
43 are probable  members of M55. This sample is dominated by
pulsating blue straggler stars but also includes 5 eclipsing binaries,  
three of which are main sequence objects.
The survey also identified several candidate blue, yellow and red 
straggler stars belonging to the cluster. We detected 15
likely members of the Sgr dSph galaxy located behind M55. 
The average PM for these stars was measured to be  
($\mu_{\alpha}cos\delta$, $\mu_{\delta}$)=($-$2.23$\pm$0.14, $-$1.83$\pm$0.24) mas/yr. 

\end{abstract}


\section{Introduction}

M55 (NGC 6809) is a metal-poor globular cluster (GC) in the Galactic halo 
(l = 9\degr, b = $-$23\degr),
discovered by Nicholas Louis de Lacaille in 1752. 
Over 70 variable stars are known in the field of the cluster
(\citealt{b3}; \citealt{b15}; \citealt{b19}; \citealt{b11}, hereafter JK10). 
About half of these are located 
in the blue straggler region of the color-magnitude diagram (CMD)
and 13 are RR Lyr type pulsators.
The cluster is close to us ($(m-M)_V$ = 13.87; \citealt{b6}) 
and is relatively sparse.  This makes it an ideal target for 
a measurement of the cluster proper motion (PM). 
Several authors have shown 
that for nearby GCs ground based CCD observations with temporal baselines 
of 3-5 years are sufficient to efficiently separate member stars from 
field interlopers (\citealt{b1}; \citealt{b23}; \citealt{b2}; \citealt{b14}).

In Sec. 2 we describe the observational data and methods used
to prepare CCD images for the analysis. The procedures employed for the
determination of relative PMs of individual stars and for the determination 
of the absolute PM of the cluster  are presented in Sec.3. 
A discussion of our results is given in Sec. 4 and this is followed
by a Summary.    

\section{Data selection and preparation}

The images of M55 analyzed in this paper were selected from the data collected
by the Cluster AgeS Experiment (CASE; \citealt{b10}) project 
during the period 1997--2008. Observations were 
made with the 2.5-m  du Pont telescope 
at Las Campanas Observatory. All images were obtained with the same 
detector and the same set of filters. We used the 2K$^2$ Tek$\#$5 CCD camera 
with a scale of 0.259\arcsec/pixel and a field of 
view of 530\arcsec$\times$530\arcsec. CASE
observed M55 as part of a program 
to obtain follow-up photometric observations of specific variables and to survey the 
central part of the cluster for detached eclipsing binaries.
For the present analysis we selected four pointings,
named F1-F4. The locations of these fields are shown in Fig. 1. The field centers
are at the following (RA, Dec) coordinates 
for Epoch 2000: F1 (294.9975\degr, $-$30.9621\degr), F2
(294.9788\degr, $-$30.9316\degr), F3 (294.9668\degr, $-$30.9499\degr) and F4
(295.0200\degr, $-$30.9707\degr). 
The set of frames selected for a given 
observing run and for a given pointing is called a `data-set'.
Each data-set includes the $V$ frames with the best seeing obtained for a given observing run.
We considered only frames with seeing better than 1.1\arcsec and 
obtained at an air-mass less than 1.1, avoiding observation obtained through clouds or with a bright sky.
A summary of the final data-sets is given in Table 1.

For each data set we constructed an averaged, 
high signal-to-noise ratio frame using the 
{\sc DIAPL}\footnote{DIAPL -- Difference Image Analysis PL {\tt
http://users.camk.edu.pl/pych/DIAPL} developed by W. Pych)} package.
In brief, individual frames were transformed to the coordinates  of a 
chosen reference image. The point spread function (PSF) of each frame 
was also transformed to match the PSF of the reference image 
(i.e.the  transformation kernel between a given frame and the reference image was derived).
The best seeing frame from a given data-set served as the reference image.
The transformed frames were subsequently stacked by averaging
to form what we later call an `average frame'.
During this procedure  the reference as well as the individual frames were divided 
into 16 overlapping sub-frames to reduce effects of  PSF variability.

We decided to obtain PMs for stars detected on the averaged frames 
of fields F1 and F2 for seasons 1997 and 1999, respectively. 
These frames are characterized by
particularly  good  seeing and high S/N. We label these the `master' 
frames. The observations of fields F1 and F2 span 
9 and 11 years, respectively.  Fields F1 and F2 show significant 
overlap 
with fields F3 and F4. Therefore, most of stars detected on F1-1997 and 
F2-1999 images are present on the remaining averaged frames. 
A master list of stars was extracted from frames F1-1997 and 
F2-1999 using the {\sc DAOPHOT/ALLSTAR} package (\citealt{b22}).
The reductions were made on 16 overlapping sub-frames 
resulting from a 4x4 division of the respective averaged frames. 
A Moffat function with linear 
spatial variability was used to characterize the PSF.
Due to crowding the photometry was extracted
in an iterative way, gradually decreasing the detection threshold. 
An effort was made to avoid artificial splitting of bright stars
which can happen when one uses an automatic procedure to detect
missed objects in star subtracted images. 
During  the last iteration the
residual images were inspected by eye to find previously undetected stars. 
The master lists  contained 47333 and 42725 objects for fields F1 and F2,
respectively. In the next step these lists were transformed to the 
coordinates of the {\it averaged frames} for different epochs. 
Profile photometry (as well 
as PSF modeling) was measured for these frames with 
ALLSTAR parameter set to $re=1$. 
In this way the positions of objects from the master lists have been re-determined 
for a given averaged frame. 
As before, we used a 4$\times$4 mosaic of sub-frames for actual reductions. 

\section{Proper motions}

\subsection{Proper motion measurements}

Relative PMs of individual stars were derived from their positions
measured at different epochs with respect to nearby cluster stars. 
This was done using a procedure
resembling that described by \citet{b1}.
As a first guess we selected as cluster stars objects located along
the main sequence and red giant and horizontal branches  on the CMD. For this purpose we constructed 
deep $V/(V-I)$ CMD's for both analyzed fields, including only stars with high
quality photometry. The photometric quality was 
judged based on CHI and SHARP fit parameters returned by {\sc ALLSTAR}. 
PM measurements were only made for stars with $V<21.0$.    

For any given star we selected a set of `grid stars' located inside an
$\sim$80\arcsec$\times$80\arcsec square,  centered on the star to be measured. 
Only likely cluster members were  considered at this step. 
The median number of grid stars was equal to 203 for F1 and 202 for F2. The grid stars
were used to calculate the local geometrical transformation between the
appropriate master frame
and a given average frame. This transformation was calculated using the  
IRAF\footnote{IRAF is distributed by the National Optical Astronomy
Observatory, which is operated by the Association of Universities for
Research in Astronomy, Inc., under a cooperative agreement with the
National Science Foundation.} $immatch.geomap$ and $immatch.geoxytran$ tasks.
We used a two-dimensional 3rd order Chebyshev polynomial as the transformation 
function. Subsequently, we calculated the expected  coordinates 
$(X_{C}, Y_{C})$ of the star
on the average frame based on its $(X_{0}, Y_{0})$ coordinates 
on the master frame.
The relative motion was  derived as the difference between the calculated and 
observed coordinates: $dX=X_{C}-X_{0}$ and $dY=Y_{C}-Y_{0}$. The shifts 
$dX$ and 
$dY$ were measured for all suitable average frames. Finally, the PM of a 
given star
was calculated by a linear fit to $dX$ and $dY$ as a function of time. 
Figure 2. shows as an example the PM measurement for star \#13300383.

Individual points were weighted by the amount of flux corresponding
to a given `average frame'. In practice this was done by summing 
fluxes measured for several bright isolated stars on individual frames 
before averaging.  We then took the inverse value of the summed flux 
to compute the relative weight for each `average frame'.
PM measurements were only attempted  for objects with positions determined on 
at least four epochs spanning at least 4 years. 
We retained only those objects for which the significance of a linear 
fit to the PM exceeded 99\%
(i.e. a $\chi^2$ test was performed with a false detection rate 
of $\alpha$ = 1.0 \%). 

PM measurements were performed in two steps. In the first step we 
made preliminary PM estimates. 
In the second step we used only reference stars for which 
reliable PMs were obtained in step one. Additionally, positions of 
reference stars  
were shifted back to the reference epoch 
(F1-1997 or F2-1999) to minimize errors due to the movement of `grid stars'. 
These new positions were used
to calculate geometrical transformations in the second iteration. 
Figure 3 shows that the total RMS of the geometrical 
transformation does not change significantly with time 
and is weakly seeing-dependent.

Equatorial coordinates for the measured stars were obtained  using 
frame astrometric solutions based on a set of stars selected from the
UCAC3 catalogue (\citealt{b24}). We considered only stars with $V<17$ which 
resulted
in 671 and 636 stars for fields F1 and F2, respectively. The average 
residual of the 
adopted solutions is 0.15 arcsec. Our catalog of M55 stars includes a total of 16954 stars 
with a measured relative PM. Table 3 presents the first few lines of the
complete catalogue which is available in electronic form at {\tt html://case.camk.edu.pl}.

\subsection{Error estimates}

Figure 4 shows the  difference in measured PMs for 11398 stars 
measured on  both fields F1 and F2.
For 66\% of these stars the difference is smaller than $\Delta\mu = 0.34$ mas/yr.
This can be adopted as a robust estimate of the average error of a PM determination for 
the whole sample. For 95\% of the stars the difference is smaller than $\Delta\mu = 0.94$ mas/yr.

Figure 5 shows the  dependence of ${\rm log_{10}}\sigma_{\mu}$ on $V$ magnitude. 
For stars 
with  $V\approx 19.0$ the median value of $\sigma_{\mu}$ is equal to 0.47 mas/yr.
At $V=20$ the median value of  $\sigma_{\mu}$ reaches 0.65 mas/yr. 
An increased scatter in $\sigma_{\mu}$ can be noted 
for $V< 14.5$. This arises  from the saturation
of bright stars on some frames resulting in fewer available 
epochs for the PM measurement. Note that the $\sigma_{\mu}$ values below $V \approx 17.5$ arise due to  
systematic effects and for fainter magnitudes are related to photon noise statistics. 
The spread of the $\sigma_{\mu}$ values may be due to the methodology
of the reductions (fitting and coordinate transformations) 
and to the number of points used in the astrometric solutions.

\subsection{Completeness}

We assessed the completeness of our PM survey as a function of $V$ magnitude and 
radial distance from the cluster center. Completeness was defined as the ratio
of the number of  stars 
with final PM measurements to the number of stars for
which we attempted to make a PM measurement.  
The results are shown in graphical
form  in Figs. 6 and 7. Only points with relative uncertainty
smaller than 20\% were plotted. 
The completeness exceeds 70\%  for $13 < V < 17$ 
and  drops to 25\% at $V=20.0$.
As expected, the completeness increases with distance from the 
cluster center, flattening at a distance of about 4 arcmin. 
The larger incompleteness observed  at the cluster center 
is due to effects associated with crowding. 
The apparent peak at $d=4$~arcmin,  evident for all magnitude ranges,
results from the fact that for more distant stars often 
only a few epochs were available. No attempt was made to 
estimate the completeness of the initial master list of stars
detected in the studied field.

\subsection{Cluster membership probabilities}

A vector point diagram (VPD) including 16945 stars with reliably 
derived relative  PMs is presented in Fig. 8. Only 2.3\% (394 objects)
of the measured objects have $\mu >$ 3.0 mas/yr and these are likely
field stars. To estimate the probability that a star is a cluster member we
use the `local-sample method' described in \citet{b17} with further
improvements proposed by \citet{b18}. In our case, however, field  stars
comprise only a very small fraction of the total sample. 

We therefore decided to assign stars to one of three groups 
based on location in the  VPD diagram and on the error of the
measured PM.
All objects with a measured PM  were divided into five sub-samples in magnitude:
$13<V<17$ (N = 1048), $17<V<18$ (N = 2511), $18<V<19$ (N = 5140), $19<V<20$
(N = 5715) and $20<V<21$ (N = 2531). For each  bin we selected stars 
with {$\mu <$ 1.8} mas/yr and  calculated for these average values and  
standard deviations of $\mu_{\alpha}cos\delta$ and $\mu_{\delta}$ 
($M_{\alpha}$, $M_{\delta}$, $S_{\alpha}$, $S_{\delta}$). The total 
value of the standard deviation of the PM was then calculated as 
$S=(S_{\alpha}^{2} + S_{\delta}^{2})^{1/2}$.
Average values 
and standard  deviations of th eindividual errors of $\mu_{\alpha}cos\delta$ 
and $\mu_{\delta}$ were also calculated 
($ME_{\alpha, \delta}$, $SE_{\alpha, \delta}$). 
Subsequently, the total values $ME=(ME_{\alpha}^{2} + ME_{\delta}^{2})^{1/2}$
and $SE=(SE_{\alpha}^{2} +  SE_{\delta}^{2})^{1/2}$ were calculated.
The results are listed in Table 2. 
The histograms showing the distribution of both components  of PM for 
the brightest and for the faintest bin in $V$ are shown in Fig. 9. 
Stars were assigned to three classes of membership. Those with  
$M > 3.0\times S$ are considered non-members (mem=0; N = 379). 
Possible members (mem=1; N = 187)
are stars which  $M \leq 3 \times S $  but with 
$\sigma_{\mu} > ME +3.0\times SE$).
Objects with $M \leq 3 \times S $ and 
$\sigma_{\mu} \leq ME +3.0\times SE$ are considered  PM 
members of the  cluster  (mem=2; N = 16379). The membership class 
defined as above is listed in 12-th column of Table 3.

\subsection{Absolute proper motions}

Among the objects remaining on the star-subtracted  averaged frames there are
a number of relatively faint, unresolved galaxies. 
These were selected by  visual inspection of both the 
{\it averaged frames} and {\it averaged frames} with stars subtracted.  
The sample includes a total of 70 galaxies from fields F1 and F2.   
Eleven of these are undoubtedly relatively bright compact galaxies. 
Due to crowding of the stellar field we decided to extract 
photometry of the galaxies simultaneously with photometry of the nearby stars 
using the {\sc GALFIT 3.0} code (\citealt{b16}).
This was done on (30$\times$30 or 60$\times$60 pixel$^{2}$) 
sub-frames centered on a given galaxy and extracted from the 
relevant averaged frame.
 
For every sub-image a PSF model was constructed using a few bright stars 
and {\sc IRAF}'s daophot routines. 
All but two  of the galaxies  were modeled with a 2-D Sersic function. The fit 
included the closest stars, some of which were present in
galaxy foreground. The sky background was also fitted. 
The coordinates of the galaxies derived this way were used 
to calculate their PMs relative to the cluster.
This was done  using the same procedure we applied
earlier to stars. 
We excluded two galaxies from the sample
since their $dX$ and $dY$ shifts showed non-linear dependence on time. 
We interpret this as due to problems with the proper 
determination of their coordinates.
The weighted average of the relative PM based on the remaining 9 galaxies is 
($\mu_{\alpha}cos\delta$, $\mu_{\delta}$) = ($-$3.19 $\pm$ 0.37, $-$9.27 $\pm$ 0.80) mas/yr. 
\citet{b25} in their search for cataclysmic variables in M55 identified 7 blue stellar-like objects. 
One of those, M55-B1, was classified as QSO based on its light curve.
The object is  present in our catalogue as 
\#23100322 with a PM value similar to the average PM of our sample of 9 galaxies. 
Another blue object, 
\#22401066  (\#17550 in the \citealt{b25} numbering of their reference frame), has a
PM indicating that it is a cluster member.
We failed to obtain PMs for the remaining 5  blue stellar-like objects
from \citet{b25} because they are too faint. 
The weighted average PM based on 9 galaxies and M55-B1 is 
($\mu_{\alpha}cos\delta$, $\mu_{\delta}$) = ($-$3.30 $\pm$ 0.20, $-$9.14 $\pm$ 0.17) mas/yr.
A VPD for these 10 objects is shown in Fig. 10. 
The point with the largest uncertainty corresponds to an edge-on galaxy. 
After dropping this object we obtained 
($\mu_{\alpha}cos\delta$, $\mu_{\delta}$) = ($-$3.31 $\pm$ 0.10, $-$9.14 $\pm$ 0.15) mas/yr. 
This value was obtained assuming ($\mu_{\alpha}^{c}cos\delta$, $\mu_{\delta}^{c}$) = (0.0,0.0) mas/yr. 
Our value can be compared with result of \citet{b4} who obtained a
PM for M55 of ($-$1.57 $\pm$ 0.62, $-$10.14$\pm$ 0.64) mas/yr. Their measurement 
was obtained using  wide-field photographic plates with a base-line of 25 years.

\section{Discussion}

\subsection{Color-magnitude diagram}
Figure 11 shows a $V/(B-V)$ CMD of M55 based on photometry from JK10.
The left panel includes about 10 thousands stars with good quality photometry.
By selecting stars classified as certain PM members 
(mem=2 in Table 3) we obtained the CMD shown 
in the right panel of Fig. 11.

\subsection{Binary stars}

A clear sequence running parallel  to the main-sequence can be seen in the cleaned CMD. We
interpret these objects to be binary stars in M55. The
binary content of M55 was  discussed at length 
by \citet{b21}. These authors used HST/ACS images
centered on the cluster core. Our data show that a population of
binaries with $q\approx 1$  is also present in the more 
external parts of M55. 
The binary sequence crosses the cluster turn-off and 
then extends to the blue including several candidate blue straggler (BS) stars. 
In addition to the BSs we note the presence of candidate 
yellow  and red straggler stars. These objects are located above the
main-sequence turnoff on both sides of the lower part of the subgiant branch.
A definitive determination of the  membership status of these 
red/yellow/blue stragglers can be made based on measurements of 
their radial velocities. M55 has a large radial velocity, 
$V_{rad}=174.8$~km/s (\citealt{b6}), and it should be easy to separate 
field stars from cluster members with high confidence. 
\subsection{Variable stars and blue stragglers}

Our PM survey can be used to assign tentative membership
for variables  and for other interesting objects from the 
cluster field. We compiled a list of variables and candidate blue
stragglers from \citet{b15}, \citet{b19}, \citet{b12} \& JK10. 
Our PM catalog includes  52 out of 71 known variables and 
46 out of 65 candidate BSs.
These stars  are listed  in Table 4. 
In Fig. 12 we present a
CMD of M55 showing the upper main-sequence and BS regions. 
All variable BSs with available PMs turned out to be probable members of the cluster. 
The BS region contains some PM members of M55 not known to be variables.
The light curves of these stars obtained by JK10 were re-examined in detail 
but  none of them showed any evidence for variability.     
Two of the variable BSs are eclipsing binaries. 
Variables which with high confidence do not belong to the cluster  
are: V15, V49, V50 and V51. In addition the BSs  BSS-7, BSS-27, BSS-31 \& BSS-39 from \citet{b12} 
are probable field stars. Individual cases should 
be checked taking into account individual PM errors. Further radial velocity 
studies of all
variable stars are needed to reliably establish their membership and
evolutionary status.%}

\subsection{Sagittarius dSph}

The variable V15 has been found by \citet{b15} to be an RR Lyr 
star belonging  to the Sagittarius dSph galaxy.
Part of this extended  galaxy is present in the background of M55 
(\citealt{b8}; \citealt{b13}). 
We measured the PM of V15 to be  
($\mu_{\alpha}cos\delta$, $\mu_{\delta}$)=(1.14 $\pm$ 0.31, 7.81 $\pm$ 0.36) mas/yr. 
The variable is located on the VPD inside an apparent clump  formed by 
a few dozen stars. We identify this group of stars as likely members 
of the Sagittarius dwarf. For 29 stars 
surrounding V15 on the VPD we  obtained a weighted mean absolute PM  
of ($\mu_{\alpha}cos\delta$, $\mu_{\delta}$)=($-$2.23$\pm$0.14, $-$1.83$\pm$0.24) mas/yr. 
This position in the VPD was calculated iteratively with the PM of V15 as a starting point 
and using stars within a circle of radius of $r=1$~mas/yr. The measured value of the average 
PM does not change beyond the quoted errors if this procedure is repeated for 
radii  of 0.75 and 1.25 mas/yr. The positions of these stars on the 
cluster CMD are shown in Fig. 11. We conclude that this subsample  
is composed
of upper main-sequence  and subgiant stars from the Sagittarius dwarf galaxy.
For a comparison, \citet{b26} determined the absolute 
PM of the Sagittarius dwarf to be  
($\mu_{\alpha}cos\delta, \mu_{\delta})=(-2.75 \pm 0.20, -1.65 \pm 0.22$) mas/yr
based on archival HST images. Their measurement was made at positions 
located 7.3 \& 9.7 deg NW from the center of M55.
Earlier determinations of the PM for the galaxy were reported by \citet{b5} and 
by \citet{b9}. 

\section{Summary}

We have performed the first proper motion study of the globular cluster M55 based on 
ground-based CCD images. Relative PMs were derived for  
16945 stars  to $V<21.0$ over a temporal baseline of up to 11 years. 
We assigned a cluster memberships status 
for all stars with available PM measurements.
The absolute PM of M55 was measured to be 
($\mu_{\alpha}cos\delta$, $\mu_{\delta}$) = ($-$3.31 $\pm$ 0.10, $-$9.14 $\pm$ 0.15) mas/yr.
The absolute PM was derived for a portion of the Sagittarius dSph 
galaxy located in the background of the cluster. 
The results of our survey allow a selection of cluster
variables, as well as cluster blue/yellow/red stragglers, with high certainty.

\section*{Acknowledgments}

KZ \& JK were supported by the Foundation for Polish Science through grant MISTRZ, 
by the grant N N203 379936 and N N203 406139 from the Ministry of Science and Higher Education.
I.B.T acknowledges the support of NSF grant AST-0507325. We thank 
Dr. F. van Leeuwen for detailed and helpful comments on an earlier version
of this paper. We thank M. Rozyczka for a careful reading of the manuscript.

\clearpage
% Table 1.
\begin{table}
\caption{Summary of M55 observations used in this study.}
\label{datasets}
\begin{tabular}{@{}lcccc}
dataset & $<$HJD-2450000.0$>$ & N & $<$exptime$>$ & $<$seeing$>$\\
        & [yr]          &   & [s]     &  [\arcsec]\\
\hline
F1-1997  &  1.65 & 12 & 60 & 0.80\\
F1-2006  & 10.70 &  6 & 75 & 1.06\\
F1-2007  & 11.86 &  8 & 40 & 1.08\\
F1-2008  & 12.66 & 21 & 35 & 0.90\\
\\
F2-1999  &  3.68 & 23 & 120 & 0.83\\
F2-2001  &  5.65 &  8 & 100 & 0.93\\
F2-2003  &  7.58 & 18 & 85 & 0.86\\
F2-2004A &  8.63 &  3 & 80 & 1.05\\
F2-2004B &  8.78 & 17 & 50 & 0.89\\
F2-2006A & 10.57 & 19 & 40 & 0.70\\
F2-2006B & 10.64 & 15 & 60 & 0.87\\
F2-2008  & 12.68 & 14 & 50 & 0.67\\
\\
F3-2006A & 10.71 &  8 & 65 & 0.82\\
F3-2006B & 10.72 &  7 & 50 & 0.83\\
F3-2007A & 11.70 &  4 & 60 & 0.95\\
F3-2007B & 11.77 & 13 & 60 & 0.78\\
F3-2008  & 12.73 & 14 & 85 & 0.89\\
\\
F4-2007  & 11.75 & 18 & 60 & 0.82\\
\hline
\end{tabular}
\end{table}

% Table 2.
\begin{table}
\caption{PM statistics for magnitude bins.
Only stars with $\mu < $ 1.8 were used. See text for details.}
\label{datasets}
\begin{tabular}{@{}ccccccc}
V & $M_{\alpha}$ & $M_{\delta}$ & $S_{\alpha}$ & $S_{\delta}$ & ME & SE\\
\hline
13--17 &  0.038 & -0.010 & 0.308 & 0.324 & 0.268 & 0.152\\
17--18 &  0.022 & -0.010 & 0.346 & 0.362 & 0.294 & 0.185\\
18--19 & -0.000 & -0.019 & 0.408 & 0.414 & 0.396 & 0.195\\
19--20 & -0.002 & -0.010 & 0.512 & 0.514 & 0.555 & 0.175\\
20--21 &  0.023 & -0.024 & 0.644 & 0.658 & 0.710 & 0.162\\
\hline
\end{tabular}
\end{table}

% Table 3.
\begin{table*}
\hspace{-8cm}
\begin{minipage}{106mm}
\caption{   
First lines of the electronically available PM catalogue. Columns: (1) star ID  (starting with 1 and 2 for
fields F1 and F2, respectively); (2) \& (3) equatorial coordinates $(\alpha,\delta)_{2000.0}$ for epochs 1997.41 and 1999.46
for F1 and F2, respectively; (4) \& (5) XY pixel coordinates on reference frames; (6)--(9) PMs and their errors; (10) number of epochs used;
(11) temporal baseline; (12) cluster membership (for explanation see Sec. 3.4);
(13) V magnitude.}
\begin{tabular}{@{}ccccccccccccc}
ID & $\alpha$ & $\delta$ & $X$ & $Y$ & $\mu_{\alpha}cos\delta$ & $\sigma_{\mu_{\alpha}cos\delta}$ & $\mu_{\delta}$ & $\sigma_{\mu_{\delta}}$ & N & dT & mem & V\\
\\
(1) & (2) & (3) & (4) & (5) & (6) & (7) & (8) & (9) & (10) & (11) & (12) & (13)\\
$[\#]$ & [\degr] & [\degr] & [pixel] & [pixel] & [mas/yr] & [mas/yr] & [mas/yr] & [mas/yr] & $[\#]$ & [yr] & [-] & [mag]\\
\\
\hline
11100007 &  295.053729 &  -30.902151 &  449.000 &  149.951 &   0.25 &   0.22 &  -0.51 &   0.29 &  8 & 10.217 & 2 & 14.586\\
11100009 &  295.058880 &  -30.928524 &  362.498 &  509.429 &  -0.66 &   0.22 &  11.84 &   0.19 &  9 & 11.033 & 0 & 14.810\\
11100020 &  295.066402 &  -30.924406 &  277.438 &  446.329 &  -0.05 &   0.17 &   0.65 &   0.21 &  5 & 11.017 & 2 & 16.792\\
11100021 &  295.060880 &  -30.928912 &  338.443 &  513.105 &  -0.36 &   0.32 &   1.24 &   0.16 &  9 & 11.025 & 2 & 16.796\\
11100023 &  295.046117 &  -30.915949 &  525.736 &  346.697 &   0.53 &   0.52 &  -0.31 &   0.79 & 13 & 10.217 & 1 & 16.980\\
11100026 &  295.071004 &  -30.916422 &  230.649 &  332.343 &   0.47 &   0.20 &   0.06 &   0.20 &  5 & 11.017 & 2 & 17.029\\
11100032 &  295.066432 &  -30.905801 &  295.049 &  189.641 &   0.09 &   0.14 &   0.15 &   0.22 &  6 & 11.017 & 2 & 17.223\\
11100039 &  295.056244 &  -30.921224 &  400.756 &  410.949 &  -0.10 &   0.11 &  -0.20 &   0.11 &  9 & 11.025 & 2 & 17.359\\
11100045 &  295.047850 &  -30.924146 &  497.279 &  458.297 &  -0.01 &   0.06 &  -0.22 &   0.14 & 11 & 11.033 & 2 & 17.441\\
11100046 &  295.070611 &  -30.926622 &  225.463 &  473.366 &  -0.61 &   0.15 &   0.88 &   0.07 &  5 & 11.017 & 2 & 17.455\\
\hline
\end{tabular}
\end{minipage}
\end{table*}

% Table 3.
\begin{table*}
\caption{Tentative membership status (mem) for variables  and for BSs from Lanzoni et al. (2007; $\rmn I\rmn D_{\rmn L}$).}
\label{variables_BSS}
\begin{tabular}{@{}rccc rccc rccc}
ID  & $\rmn I\rmn D_{\rmn L}$ & $\rmn I\rmn D_{\rmn P\rmn M}$ & mem & ID  & $\rmn I\rmn D_{\rmn L}$ & $\rmn I\rmn D_{\rmn P\rmn M}$ & mem & ID  & $\rmn I\rmn D_{\rmn L}$ & $\rmn I\rmn D_{\rmn P\rmn M}$ & mem \\
\hline
V02 & --- & 24300001 & 2 &       V32 & BSS-21 & 22400204 & 2 &   V62 & --- & 23201072 & 2\\
V04 & --- & 12200014 & 2 &       V33 & BSS-51 & 13400039 & 2 &   V63 & BSS-35 & 23300072 & 2\\
V05 & --- & 22405193 & 2 &       V34 & BSS-61 & 12400077 & 2 &   V64 & --- & 23300099 & 2\\
V06 & --- & 12200010 & 2 &       V35 & BSS-52 & 23200027 & 2 &   V65 & --- & 13300228 & 2\\
V07 & --- & 12220322 & 2 &       V36 & BSS-53 & 23300087 & 2 &   V67 & --- & 14400131 & 2\\
V08 & --- & 22400024 & 2 &       V37 & BSS-55 & 13300163 & 2 &   V69 & BSS-45 & 13400057 & 2\\
V10 & --- & 12100004 & 2 &       V38 & BSS-20 & 13300141 & 2 &   --- & BSS-01 & 22300208 & 2\\
V11 & --- & 21300008 & 2 &       V40 & BSS-60 & 12200189 & 2 &   --- & BSS-02 & 13200131 & 2\\
V12 & --- & 12300018 & 2 &       V41 & BSS-10 & 22400135 & 2 &   --- & BSS-07 & 22300198 & 0\\
V15 & --- & 14400150 & 0 &       V42 & BSS-25 & 13200170 & 2 &   --- & BSS-08 & 12300182 & 2\\
V16 & BSS-56 & 12200151 & 2 &   V44 & --- & 21400232 & 2 &       --- & BSS-14 & 23400059 & 2\\
V17 & BSS-58 & 11300070 & 2 &   V45 & BSS-32 & 12200109 & 2 &   --- & BSS-15 & 22300189 & 2\\
V18 & BSS-22 & 21300093 & 2 &   V47 & BSS-06 & 22300183 & 2 &   --- & BSS-16 & 23300108 & 2\\
V19 & BSS-12 & 22300308 & 2 &   V48 & BSS-03 & 12200099 & 2 &   --- & BSS-19 & 22400262 & 2\\
V20 & BSS-23 & 23400089 & 2 &   V49 & --- & 11300622 & 0 &       --- & BSS-27 & 13400030 & 0\\
V21 & BSS-26 & 22400085 & 2 &   V50 & --- & 21400104 & 0 &       --- & BSS-31 & 21400069 & 0\\
V22 & BSS-54 & 12400055 & 2 &   V51 & --- & 21200261 & 0 &       --- & BSS-34 & 22200033 & 2\\
V23 & BSS-62 & 13200171 & 2 &   V53 & BSS-37 & 12300123 & 2 &   --- & BSS-36 & 11300049 & 2\\
V24 & BSS-57 & 23300107 & 2 &   V54 & --- & 24300158 & 2 &       --- & BSS-39 & 23100015 & 1\\
V25 & BSS-50 & 23300058 & 2 &   V55 & BSS-46 & 21300126 & 2 &   --- & BSS-41 & 22203014 & 2\\
V26 & BSS-63 & 23300063 & 2 &   V57 & BSS-04 & 22300166 & 2 &   --- & BSS-49 & 22200293 & 2\\
V27 & BSS-24 & 23400103 & 2 &   V60 & --- & 22300197 & 2 &        & & & \\
V31 & BSS-11 & 12300229 & 2 &   V61 & --- & 12300228 & 2 &        & & & \\
\hline
\end{tabular}
\end{table*}

%%%%%%%%%%%%%%%%%%%%%

\begin{figure*}
\begin{center}
   \leavevmode
   \epsfxsize=90mm
   \epsfbox{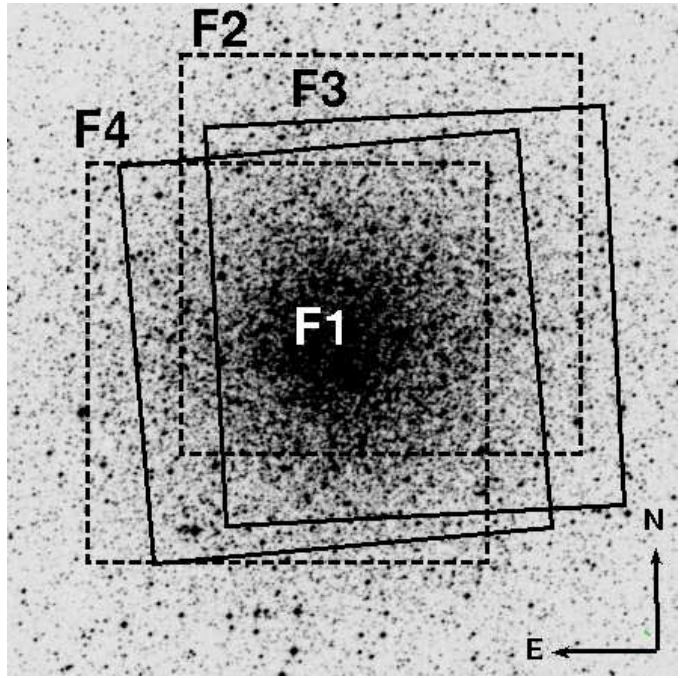}
\end{center}
\caption{Positions of the F1, F2, F3, F4 fields overlaid on a
DSS R-band  image of M55. The field of view is 
15\arcmin$\times$15\arcmin.}
\label{fields}
\end{figure*}

\begin{figure*}
\begin{center}
   \leavevmode
   \epsfxsize=90mm
   \epsfbox{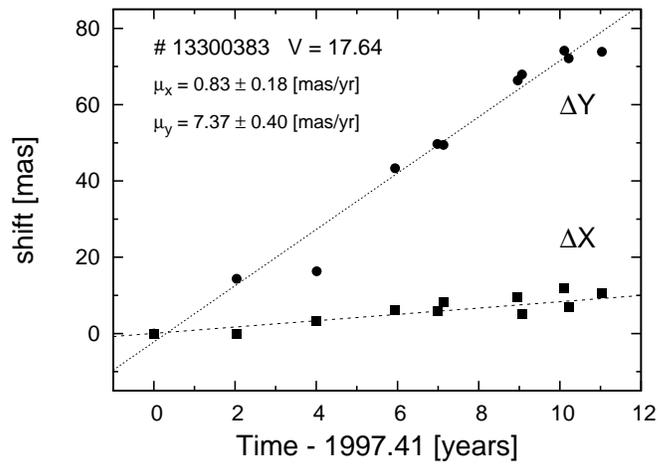}
\end{center}
\caption{
An example of PM determination for one of the  fast moving stars.
}
\label{fields}
\end{figure*}

\begin{figure*}
\begin{center}
   \leavevmode
   \epsfxsize=90mm
   \epsfbox{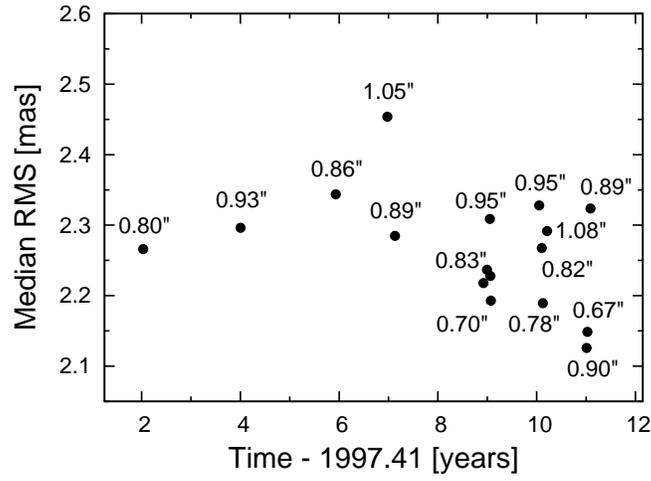}
\end{center}
\caption{ RMS of geometrical transformation
versus time for field F1. Average seeing for a given frame is indicated.}
\label{fields}
\end{figure*}

\begin{figure*}
\begin{center}
   \leavevmode
   \epsfxsize=90mm
   \epsfbox{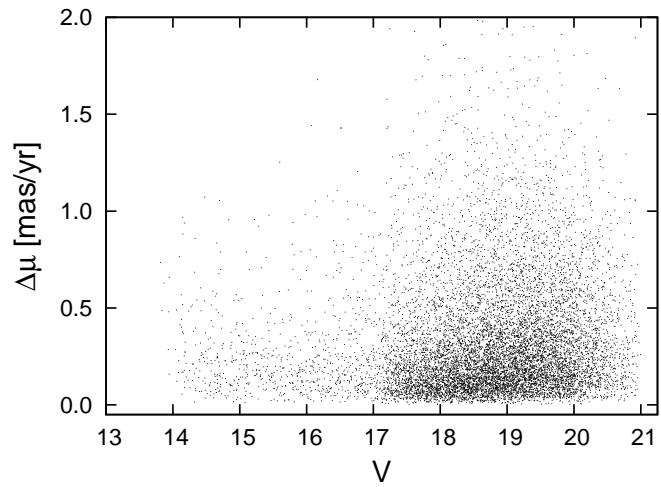}
\end{center}
\caption{Differences of PMs measured for stars in common to fields 
F1 \& F2  as a function of $V$ magnitude.}
\label{fields}
\end{figure*}

\begin{figure*}
\begin{center}
   \leavevmode
   \epsfxsize=90mm
   \epsfbox{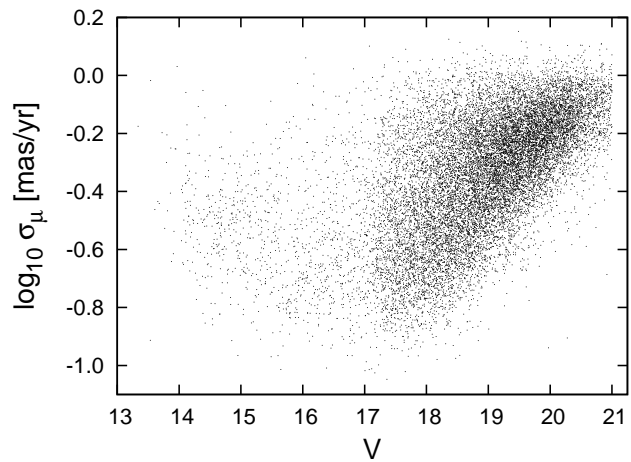}
\end{center}
\caption{PM errors.}
\label{fields}
\end{figure*}

\begin{figure*}
\begin{center}
   \leavevmode
   \epsfxsize=90mm
   \epsfbox{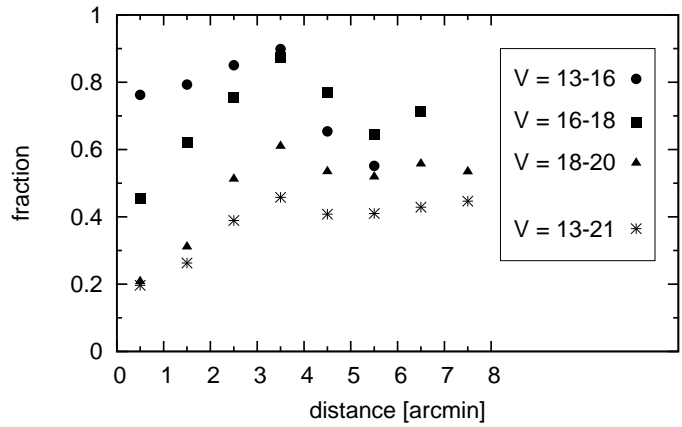}
\end{center}
\caption{Completeness of the PM catalog as a function of projected distance from the cluster center.}
\label{fields}
\end{figure*}

\begin{figure*}
\begin{center}
   \leavevmode
   \epsfxsize=90mm
   \epsfbox{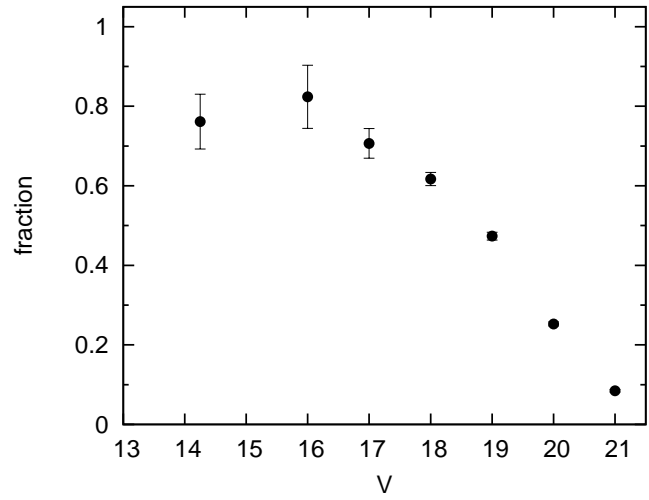}
\end{center}
\caption{Completeness of the PM catalog as a function of $V$ magnitude.}
\label{fields}
\end{figure*}

\begin{figure*}
\begin{center}
   \leavevmode
   \epsfxsize=90mm
   \epsfbox{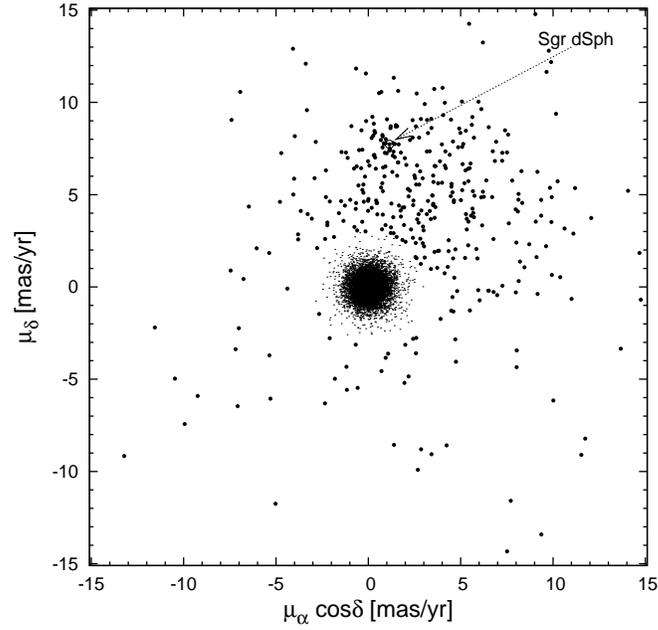}
\end{center}
\caption{The proper motion Vector Point Diagram for 16945 stars from  M55. 
The catalog includes 15 high velocity stars located outside of the
plotted area.  
Stars with  $\mu >$3.0 are marked with large dots (394 objects). 
The arrow points to the  clump of stars presumably associated with 
the Sgr dSph galaxy.}
\label{fields}
\end{figure*}

\begin{figure*}
\begin{center}
   \leavevmode
   \epsfxsize=120mm
   \epsfbox{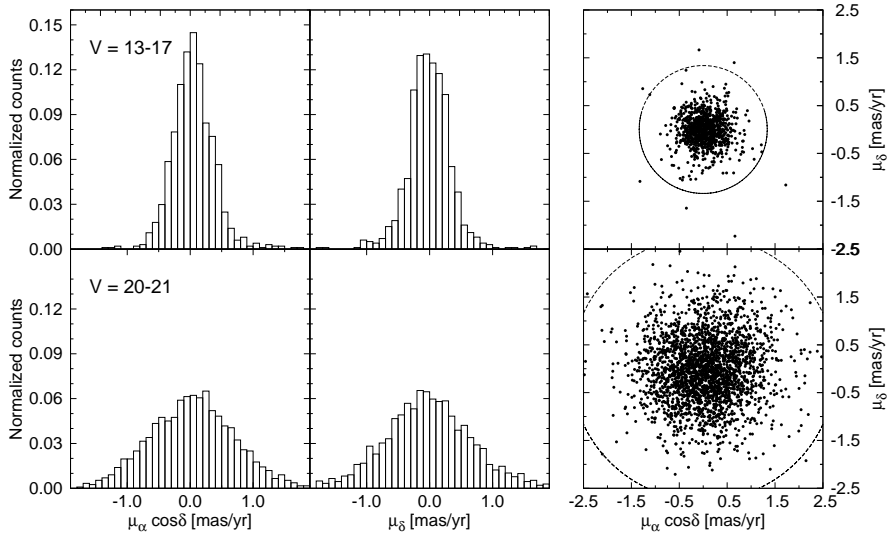}
\end{center}
\caption{The distribution 
in $\mu_{cos\delta}$ and $\mu_{\delta}$ for magnitude bins 
V = 13-17 and V = 20-21. The right panel shows a VPD for stars with 
 PM $<$ 1.8 mas/yr. Circles indicate 3$\times$S which is defined in Sec. 3.4.}
\label{fields}
\end{figure*}

\begin{figure*}
\begin{center}
   \leavevmode
   \epsfxsize=90mm
   \epsfbox{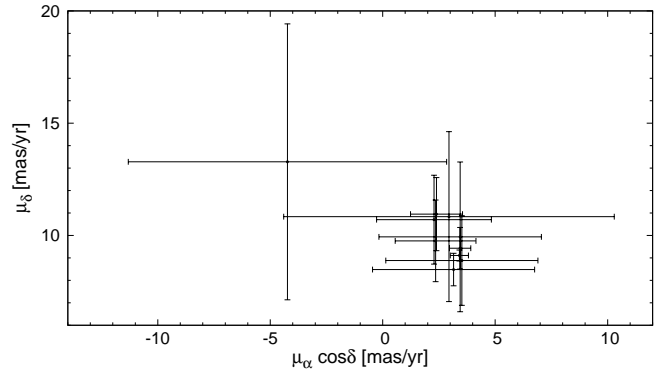}
\end{center}
\caption{Relative PMs of 9 compact galaxies and the QSO M55-B1 
measured with respect to M55 stars.}
\label{fields}
\end{figure*}

\begin{figure*}
\begin{center}
   \leavevmode
   \epsfxsize=120mm
   \epsfbox{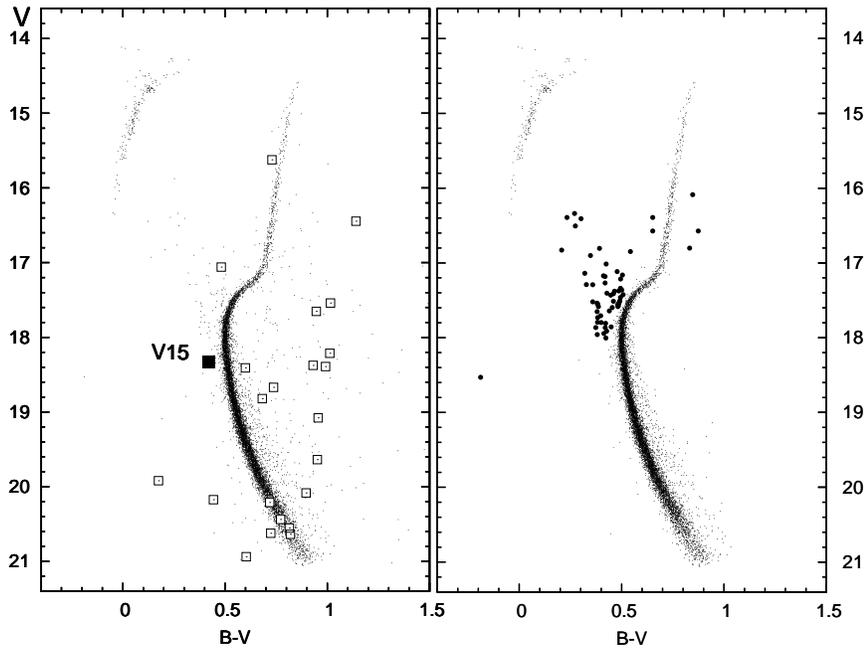}
\end{center}
\caption{
Left panel - $V/(B-V)$ CMD of M55 based on JK10 photometry. 
Stars selected as likely PM members of the Sgr dSph galaxy
are marked with squares (see Sec. 4.3).; Right panel - the CMD including 
only probable cluster members. Candidate blue/yellow/red stragglers are marked with large dots.}
\label{fields}
\end{figure*}

\begin{figure*}
\begin{center}
   \leavevmode
   \epsfxsize=90mm
   \epsfbox{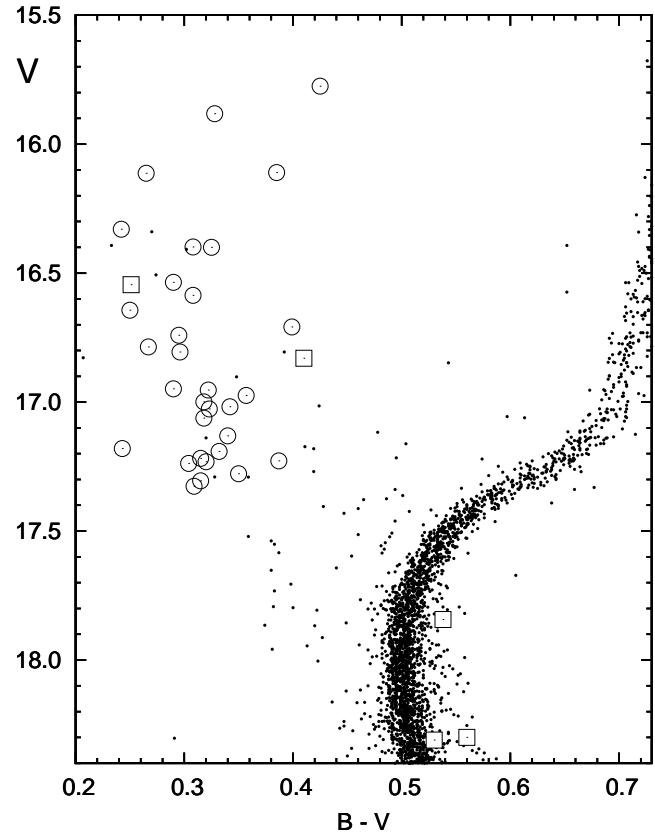}
\end{center}
\caption{$V/(B-V)$ CMD showing 
the turnoff and BS regions of M55. 
Only likely cluster members are plotted.  Circles - pulsating variables of SX~Phe type; 
squares - eclipsing binaries.}
\label{fields}
\end{figure*}

\bsp

\label{lastpage}

\end{document}